\documentclass[%
	aip,
	amsmath,amssymb,
	preprint,
]{revtex4-1}

\usepackage{amsmath, amsthm, amssymb, pifont, wasysym}
\usepackage[dvipdfmx]{graphicx}
\usepackage{bm} 
\usepackage{pifont}
\usepackage{dcolumn} 
\usepackage{color}
\definecolor{mygray}{gray}{0.5}

\newcommand {\EC}{EuCd$_2$}
\newcommand {\EZ}{EuZn$_2$}
\newcommand {\AO}{Al$_2$O$_3$}
\newcommand {\AB}{AlB$_2$}
\newcommand {\CC}{CeCu$_2$}

\newcommand {\Bout}{$B_\text{out}$}

\newcommand {\rxx}{$\rho_{xx}$}
\newcommand {\ryx}{$\rho_{yx}$}
\newcommand {\OHE}{$\rho_{yx}^\text{O}$}
\newcommand {\AHE}{$\rho_{yx}^\text{A}$}
\newcommand {\THE}{$\rho_{yx}^\text{T}$}

\newcommand {\TN}{$T_\text{N}$}

\begin{document}
	\title{Distinct topological Hall responses in {\CC}-type {\EZ} and {\EC} films}
	\author{Yuto Watanabe}
	\affiliation{Department of Physics, Tokyo Institute of Technology, Tokyo 152-8551, Japan}
	\author{Shinichi Nishihaya}
	\affiliation{Department of Physics, Tokyo Institute of Technology, Tokyo 152-8551, Japan}
	\author{Markus Kriener}
	\affiliation{RIKEN Center for Emergent Matter Science (CEMS), Wako 351-0198, Japan}
	\author{Ayano Nakamura}
	\affiliation{Department of Physics, Tokyo Institute of Technology, Tokyo 152-8551, Japan}
	\author{Masaki Uchida}
	\email[Author to whom correspondence should be addressed: ]{m.uchida@phys.titech.ac.jp}
	\affiliation{Department of Physics, Tokyo Institute of Technology, Tokyo 152-8551, Japan}
	
	\begin{abstract}
	\color{black}Rare earth intermetallic compounds crystallized in {\AB}-type and its low-symmetry derivative {\CC}-type structures potentially host diverse frustrated magnetic structures and rich magnetotransport phenomena. \color{black}We report the film growth of {\CC}-type {\EZ} by molecular beam epitaxy and the observation of topological Hall responses highly contrastive to isostructural {\EC}. While their magnetization curves are rather similar, the topological Hall effect observed in {\EZ} is simpler, with the only one component enhanced at the magnetic transition field. {\EZ} may be a unique system for studying the magnetic domain boundary effect on topological Hall responses among the {\CC}-type rare-earth intermetallic compounds.
	\end{abstract}

	\maketitle

	Transverse motion of carriers neither proportional to the magnetic field nor the magnetization has attracted broad interest both in fundamental and applied physics \cite{AHEreview}. Such a nonmonotonic Hall response originates in a change in the real-space or momentum-space Berry curvature in the magnetization process. In particular, the Hall response reflecting a real-space spin texture is termed topological Hall effect (THE), where a noncoplanar spin texture with finite scalar spin chirality, defined by the solid angle spanned by three adjacent spins, generates an emergent magnetic field on charged carriers \cite{THE,Nd2Mo2O7}. A magnetic skyrmion lattice is a typical example of the noncoplanar spin texture. Recently, the range of skyrmion host compounds has been expanded from noncentrosymmetric chiral to centrosymmetric frustrated magnets \cite{MnSi,MnGe,FeGe,GdPdSi,EuAl4}.

	As centrosymmetric magnets which can host a skyrmion lattice, {\AB}-type compounds represented by Gd$_2$PdSi$_3$ have attracted growing attention \cite{GdPdSi}. For example in Gd$_2$PdSi$_3$, a magnetic Gd triangular lattice layer and a Pd/Si honeycomb lattice layer are alternately stacked to form a two-dimensionally frustrated system, where a giant topological Hall response is observed with rather monotonic change in magnetization \cite{GdPdSi}. The {\AB}-type structure is also famous for many lower-symmetry derivatives including the CeCu$_2$-type one \cite{AlB2_derivative}. In rare-earth intermetallic compounds $RA_2$ ($R = $ La -- Lu, $A = $ Zn, Cd, or Cu) with the CeCu$_2$-type structure, a distorted $R$ triangular lattice layer and a three-dimensionally buckled $A$ honeycomb lattice layer are alternately stacked \cite{CeCu2-type}. Differences in distortion and buckling as well as magnetic anisotropy of rare-earth magnetic moments can bring about diverse frustrated magnetic structures and magnetotransport in $RA_2$. In addition to helical magnetic ordering with various propagation directions and rotation planes as observed in GdCu$_2$ and PrZn$_2$ \cite{GdCu2_mag, PrZn2_mag, DyZn2_mag}, rich topological Hall responses have been recently found for {\EC} \cite{EuCd2_THE}. However, comprehensive research on the magnetotransport in $RA_2$ is still lacking.

\begin{figure*}
		\begin{center}
			\includegraphics*[width=17cm]{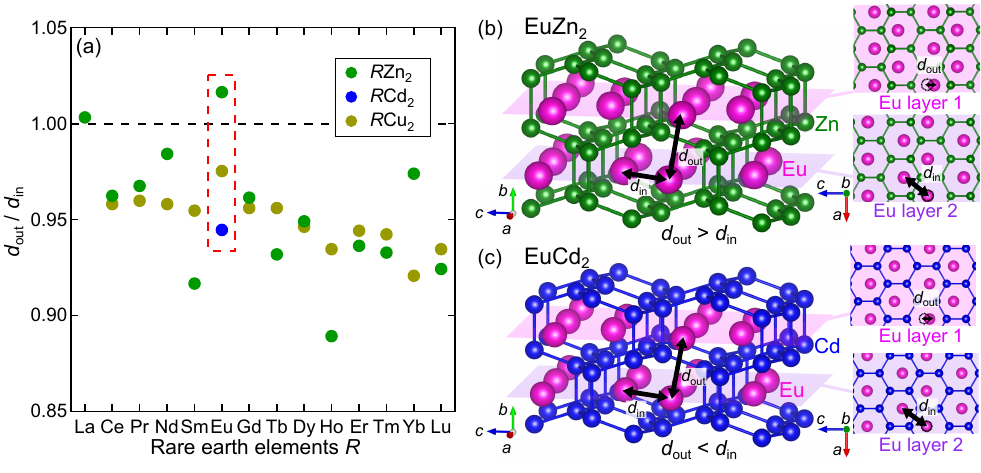}	
	    	\caption{
		Crystal structure and atomic distances of {\CC}-type rare-earth intermetallic compounds. (a) Ratio of out-of-plane and in-plane rare-earth atomic distances ($d_\text{out} / d_\text{in}$) in {\CC}-type $RA_2$ ($R = $ La -- Lu, $A = $ Zn, Cd, or Cu) \cite{EuCd2, EuZn2, CeZn2_mag, PrZn2_mag, NdZn2, NdZn2_mag, DyZn2_mag, HoZn2, YbZn2, CeCu2, GdCu2, NdCu2, EuCu2, DyCu2, HoCu2, ErCu2, TmCu2, YbCu2}. Crystal structure of (b) {\EZ} and (c) {\EC}, composed of alternate stacking of a distorted Eu triangular lattice layer and a buckled Zn or Cd honeycomb lattice layer along the $b$-axis. While pairs of the nearest neighbor Eu atoms are formed within the triangular lattice layers in {\EZ} ($d_\text{out} > d_\text{in}$), they are between the adjacent layers in {\EC} ($d_\text{out} < d_\text{in}$).
			}
	   		\label{fig1}
		\end{center}
	\end{figure*}

	Figure \ref{fig1}(a) summarizes the ratio of out-of-plane and in-plane rare-earth atomic distances ($d_{\mathrm{out}}$ and $d_{\mathrm{in}}$) in {\CC}-type intermetallic compounds \cite{EuCd2, EuZn2, CeZn2_mag, PrZn2_mag, NdZn2, NdZn2_mag, DyZn2_mag, HoZn2, YbZn2, CeCu2, GdCu2, NdCu2, EuCu2, DyCu2, HoCu2, ErCu2, TmCu2, YbCu2}. In most of them including {\EC}, pairs of the nearest neighbor rare-earth atoms are formed between the adjacent triangular lattice layers ($d_{\mathrm{out}}<d_{\mathrm{in}}$). Actually, {\EZ} is the only magnetic compound with such pairs within the triangular lattice layers ($d_{\mathrm{out}}>d_{\mathrm{in}}$), since LaZn$_2$ is nonmagnetic. Drastic changes in magnetotransport can be thus expected reflecting the different frustration arrangements of the Eu$^{2+}$ isotropic magnetic moments \cite{EuCd2, EuZn2}, as compared in Figs. \ref{fig1}(b) and \ref{fig1}(c). In this paper, we report film growth of {\EZ} by molecular beam epitaxy and observation of topological Hall responses highly contrastive to {\EC}.

\begin{figure}
		\begin{center}
			\includegraphics*[width=13cm]{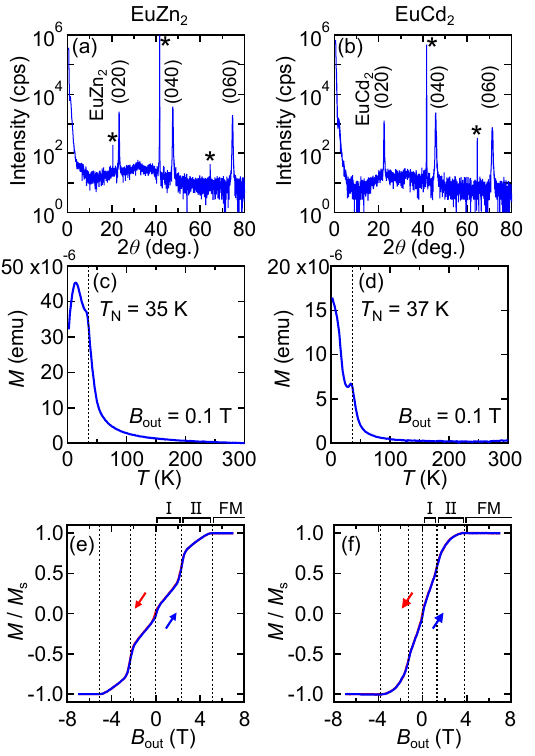}
	    	\caption{
		Magnetic properties of {\EZ} and {\EC} thin films. X-ray diffraction 2$\theta$-$\omega$ scans of (010)-oriented (a) {\EZ} and (b) {\EC} films grown on the {\AO} (0001) substrate. An asterisk denotes the diffraction peaks from the {\AO} substrate. Temperature dependence of the magnetization in (c) {\EZ} and (d) {\EC} films, measured in an out-of-plane magnetic field of $B_{\mathrm{out}}=0.1$ T. Magnetization curves of (e) {\EZ} and (f) {\EC} films taken with sweeping $B_{\mathrm{out}}$ at 2 K. Data are normalized by the saturation magnetization $M_{\mathrm{s}}$.
			}
	   		\label{fig2}
		\end{center}
	\end{figure}

	Thin films of {\EZ} and {\EC} were grown on {\AO} (0001) substrates using an Epiquest RC1100 II-V molecular beam epitaxy chamber \cite{EuAs, EuCd2_THE, EuCd2Sb2_film, EuCd2As2_film, EuCdSb2_film, EuSb2_film}. The growth temperature for {\EZ} films was typically set to 400 $^\circ$C. Eu and Zn were supplied by an effusion cell with a Zn-rich flux ratio ($P_{\mathrm{Zn}}/P_{\mathrm{Eu}} = 6$). This flux ratio is lower than in the growth of {\EC} films ($P_{\mathrm{Cd}}/P_{\mathrm{Eu}} = 15-44$) \cite{EuCd2_THE}, consistent with the less volatile nature of Zn compared to Cd. The film thickness is designed at 65 nm. As confirmed in Figs. \ref{fig2}(a) and \ref{fig2}(b), x-ray diffraction $2\theta$-$\omega$ scans for both {\EZ} and {\EC} films exhibit sharp peaks from the (020) lattice plane without any impurity peaks, indicating that the distorted Eu triangular lattice layers are stacked on the {\AO} (0001) plane.
	
	Figures \ref{fig2}(c) -- \ref{fig2}(f) compare fundamental magnetic properties taken for {\EZ} and {\EC} films. The temperature dependence of the magnetization $M$ exhibits a cusp at the N\'{e}el temperature $T_\text{N}$ = 35 K for {\EZ} and 37 K for {\EC}, almost consistent with previous reports in polycrystalline bulks \cite{EuZn2, EuCd2}. Both in the {\EZ} and {\EC} films, $M$ keeps increasing below $T_\text{N}$, suggesting that their magnetic ground states are not a simple collinear antiferromagnetic ordering. Magnetization curves, taken with sweeping the out-of-plane field {\Bout} at 2 K, are rather monotonic and also quite similar between {\EZ} and {\EC}. A transition from phase I to II occurs at the metamagnetic transition field $B_\text{m}$ = 2.3 T for {\EZ} and 1.3 T for {\EC} with tiny hysteresis, and then the Eu$^{2+}$ magnetic moments are forcedly aligned along the out-of-plane direction above the saturation field $B_\text{s}$ = 5.1 T for {\EZ} and 3.8 T for {\EC}. Another small hysteresis loop is confirmed centered at 0 T, indicating that there is a weak ferromagnetic component at the ground state of {\EZ} as well as {\EC}.
	
\begin{figure}
		\begin{center}
			\includegraphics*[width=14cm]{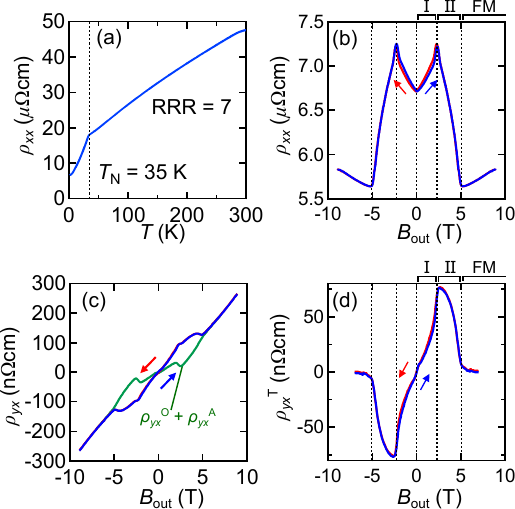}
	    	\caption{
		Magnetotransport of the {\EZ} film. (a) Temperature dependence of longitudinal resistivity $\rho_{xx}$. Out-of-plane field dependence of (b) $\rho_{xx}$ and (c) Hall resistivity $\rho_{yx}$ at 2 K. $\rho_{yx}$ data above the saturation field is fitted by a green curve consisting of the ordinary Hall component ($\rho_{yx}^{\mathrm{O}} = R_0${\Bout}) and the anomalous Hall component ($\rho_{yx}^{\mathrm{A}} = r_{\mathrm{s}}\rho_{xx}^2 M)$ with field-independent constants $R_0$ and $r_{\mathrm{s}}$. (d) Topological Hall component {\THE} at 2 K, obtained by subtracting $\rho_{yx}^{\mathrm{O}} + \rho_{yx}^{\mathrm{A}} $ from $\rho_{yx}$ in (c).
			}
	   		\label{fig3}
		\end{center}
	\end{figure}

	Figure \ref{fig3} summarizes magnetotransport observed for the {\EZ} film. As shown in Fig. \ref{fig3}(a), the temperature dependence of the longitudinal resistivity {\rxx} is metallic with a residual resistivity ratio (RRR) of 7. It exhibits a clear kink at {\TN}, reflecting strong coupling between the itinerant carriers and the localized Eu$^{2+}$ magnetic moments. In the out-of-plane field dependence shown in Fig. \ref{fig3}(b), {\rxx} exhibits a peak at $B_\text{m}$ and then negative magnetoresistance up to $B_\text{s}$. Figure \ref{fig3}(c) shows the Hall resistivity {\ryx} measured with sweeping the out-of-plane field {\Bout} at 2 K. \color{black}The hole density estimated by fitting the {\ryx} curve above the saturation field at 2 K is $1.7\times10^{22}$ cm$^{-3}$, which is similar to that of {\EC}\cite{EuCd2_THE}. \color{black}It is clear that there is a nonmonotnic Hall component in addition to the ordinary Hall resistivity {\OHE} and the anomalous Hall resistivity {\AHE}. Here {\OHE} is expressed by {\OHE}  = $R_0${\Bout} with the Hall coefficient $R_0$, and {\AHE} is extracted by {\AHE} = $r_s$\rxx$^2M$ with $r_s$ determined as a fitting parameter, since the longitudinal conductivity converted from {\rxx} is located in the so-called intrinsic region of the conventional scaling plot \cite{AHE_scaling}. Figure \ref{fig3}(d) displays the additional component obtained by subtracting {\OHE} and {\AHE} from the measured Hall resistivity. Hereafter we call this {\THE}, which is ascribed to the topological Hall component originating in frustrated spin configuration, as discussed below.

	{\THE} appears below $T_{\mathrm{N}}$, indicating that it is related to the magnetic ordering. As reported for some magnetic compounds, {\ryx} may be affected by the change in {\rxx} through a field dependence of the scattering time. In ErB$_4$, for example, a nonmonotonic Hall component has been explained by a modified two-band model incorporating this effect \cite{ErB4}. In the present case, however, the fitting curve of {\OHE}+{\AHE} swells rather in the direction opposite to the hump around 2.5 T, and thus the change in the scattering time would not be the cause.
	
\begin{figure}
		\begin{center}
			\includegraphics*[width=14cm]{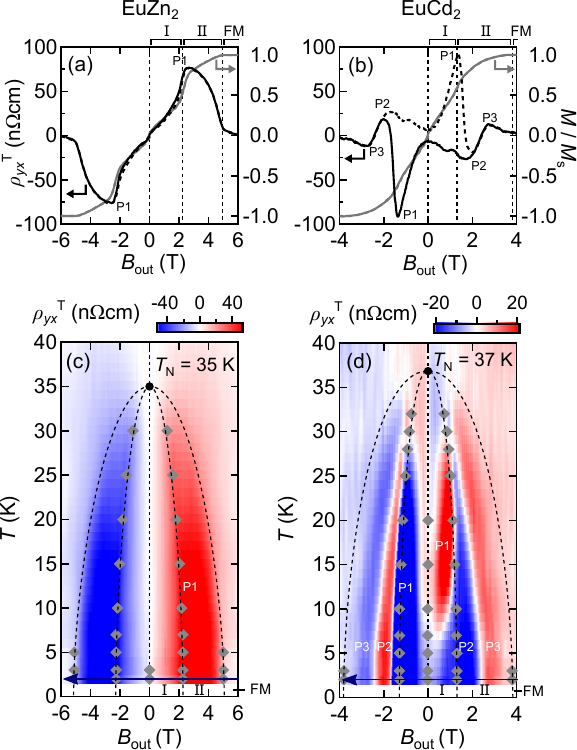}
	    	\caption{
		Comparison of the topological Hall effect observed in {\EZ} and {\EC}. Topological Hall resistivity {\THE} plotted together with the magnetization $M$ for (a) {\EZ} and (b) {\EC} films at 2 K. A solid (dashed) curve represents the data taken upon decreasing (increasing) the field from +7 (-7) T to -7 (+7) T. Mapping of {\THE} on the {\Bout}-$T$ phase diagram for (c) {\EZ} and (d) {\EC} films. Data taken in the downward field sweep is shown on the map. Magnetic transition fields determined from $dM/dB$ are denoted by a diamond, and magnetic phase boundaries are represented by a dashed curve for a guide to the eye.
			}
	   		\label{fig4}
		\end{center}
	\end{figure}

	Here comparison with {\EC} is helpful for discussion. Figures \ref{fig4}(a) and \ref{fig4}(b) compare {\THE} taken for {\EZ} and {\EC} at 2 K. In {\EZ}, only the peak P1 appears centered at $B_\text{s}$ between phase I and II. In {\EC}, on the other hand, {\THE} is characterized by three peaks P1-P3, and among them, peak P1 similarly appears centered at $B_\text{s}$ as in {\EZ}. Another difference is that a large hysteresis loop is observed in {\EC} accompanied with a sharp enhancement of P1 at low temperatures. The origin of THE in {\EC} has been understood the real-space spin Berry curvature, or noncoplanar spin textures realized on the distorted $Eu^{2+}$ triangular lattice layers \cite{EuCd2_THE}. \color{black}It has been clarified based on the temperature dependence measurements that the sign of the THE peaks P1-P3 originating from the real-space spin Berry curvature remain unchanged, while {\AHE} reflecting the momentum-space Berry curvature exhibits a sign change at 20 K. \color{black}Although the sign change of {\AHE} does not occur in {\EZ} at elevated temperatures, it is reasonable to \color{black}suggest \color{black}that THE in {\EZ} also originates in noncoplanar spin textures in the magnetization process, \color{black}considering the similarity of the magnetization properties between the two materials. \color{black} In fact, the distorted Eu triangular lattice in the $ac$-plane of {\CC}-type structure can host uncanceled local Dzyaloshinskii-Moriya interaction which favors spin configuration canted from the $ac$-plane to the out-of-plane $b$-axis, possibly leading to a noncoplanar ordering. \color{black}
	
	Figures \ref{fig4}(c) and \ref{fig4}(d) compare a color-map of {\THE} taken in the downward field sweep on the {\Bout}-$T$ phase diagram. In {\EZ}, {\THE} consisting only of the peak P1 appears antisymmetric to the magnetic field and continues to reach a maximum value at the magnetic transition field between phases I and II even upon increasing temperature. In {\EC}, the appearance of {\THE} is more complex and not antisymmetric to the field due to the large hysteresis at low temperatures, but the enhancement of P1 at the magnetic transition field is commonly confirmed in the negative field region. As the peak P1 is enhanced not within a specific phase but at a phase boundary, it is naturally ascribed to the magnetic domain boundary formed between the two phases. In evidence, it has been demonstrated that P1 in {\EC} strongly depends on the field cooling processes, which can effectively modulate the domain boundary density \cite{EuCd2_THE}. On the other hand, the differences in the appearance of THE between {\EZ} and {\EC} are probably derived from different pairs of the nearest neighbor Eu atoms and dominant magnetic interactions as illustrated in Fig. \ref{fig4}(c). While spin modulation propagating along the $c$-axis has been reported for other intermetallic compounds $R$Zn$_2$ with $d_{\mathrm{out}}<d_{\mathrm{in}}$ \cite{CeZn2_mag, PrZn2_mag, NdZn2_mag, DyZn2_mag}, distinctive magnetic structures \color{black}are \color{black}realized in {\EZ} with $d_{\mathrm{out}}>d_{\mathrm{in}}$, which deserve clarification in future studies. The appearance of a simple single-component THE also suggests that {\EZ} \color{black}is potentially \color{black}a unique system for further studying the magnetic domain boundary effect on THE among the {\CC}-type rare-earth intermetallic compounds.
		
	In summary, we have grown {\EZ} thin films by molecular beam epitaxy and revealed topological Hall responses which are in sharp contrast to {\EC}. While their magnetization curves in the magnetization process are rather similar, the topological Hall effect observed in {\EZ} is simpler, with the only one component enhanced at the metamagnetic transition field. \color{black}Hence, we propose {\EZ} as \color{black}a unique system for further studying the magnetic domain boundary effect on THE among the {\CC}-type rare-earth intermetallic compounds. 

\begin{acknowledgments}	
	We thank M. Kawasaki for help in a part of the magnetization measurements. We also thank H. Ishizuka, Y. Yamasaki, T. Nakajima, F. Kagawa, and K. Matsuura for fruitful discussions. This work was supported by JST FOREST Program Grant No. JPMJFR202N, Japan, by Grant-in-Aids for Scientific Research JP22H04471, JP22H04501, JP22K18967, JP22K20353, JP23K13666, JP24H01614, and JP24H01654 from MEXT, Japan, and by STAR Award funded by the Tokyo Tech Fund, Japan.
\end{acknowledgments}

\section*{Conflict of Interest}
The authors have no conflicts to disclose.

\section*{Author Contributions}
\textbf{Yuto Watanabe}: Data curation (lead); Formal analysis (equal); Investigation (lead); Methodology (equal); Visualization (lead); Writing - original draft preparation (lead). 
\textbf{Shinichi Nishihaya}: Formal analysis (equal), Investigation (equal), Methodology (equal), Writing-review \& editing (equal).
\textbf{Markus Kriener}: Formal analysis (supporting), Investigation (equal), Writing - review \& editing (equal).
\textbf{Ayano Nakamura}: Formal analysis (supporting), Investigation (supporting), Writing - review \& editing (equal).
\textbf{Masaki Uchida}: Conceptualization (lead); Funding acquisition (lead); Methodology (lead); Project administration (lead); Supervision (lead); Writing - review \& editing (lead).

\section*{Data Availability}
The data that supports the findings of this study are available from the corresponding author upon reasonable request.

	\clearpage

\end{document}